\documentclass[epj]{svjour}

\smartqed  

\RequirePackage{mathptmx}      
\usepackage{flushend}
\usepackage[numbers,sort&compress]{natbib}
\usepackage[colorlinks,citecolor=blue,urlcolor=blue,linkcolor=blue]{hyperref}
\usepackage{graphicx}
\usepackage{subfig}
\usepackage{dcolumn}
\usepackage{bm}
\usepackage{float}
\usepackage{xcolor}
\usepackage{amssymb}
\usepackage{amsmath}
\usepackage{amsmath,bm}
\usepackage{graphics}
\usepackage{graphicx}
\usepackage{color}
\journalname{Eur. Phys. J. C}
\DeclareMathAlphabet{\mathcal}{OMS}{cmsy}{m}{n}
\DeclareSymbolFont{largesymbols}{OMX}{cmex}{m}{n}

\usepackage{array}
\makeatletter
\newcommand{\thickhline}{%
    \noalign {\ifnum 0=`}\fi \hrule height 1pt
    \futurelet \reserved@a \@xhline
}
\newcolumntype{"}{@{\hskip\tabcolsep\vrule width 1pt\hskip\tabcolsep}}
\makeatother

\begin{document}

\title{The constraint ability of Hubble parameter by gravitational wave standard sirens on cosmological parameters}

\author{Tong-Jie Zhang\inst{1,2}\thanks{Corresponding author. Email: tjzhang@bnu.edu.cn}
        \and
        Yang Liu\inst{1}
        \and
        Zhi-E Liu\inst{3}
        \and
        Hao-Yi Wan\inst{1,4} 
        \and
        Ting-Ting Zhang\inst{5}
        \and
        Bao-Quan Wang\inst{6}
}


\institute{Department of Astronomy, Beijing Normal University, Beijing 100875, China\label{addr1}
          \and
          Institute for Astronomical Science, Dezhou University, Dezhou 253023, China\label{addr2}
          \and
          College of Physics and Electronic Engineering, Qilu Normal University, Jinan 250200, China\label{addr3}
          \and
          National Astronomical Observatories, Chinese Academy of Sciences, Beijing 100012, China\label{addr4}
          \and
          PLA Army Engineering University, Nanjing 210017, China\label{addr5}
          \and
          School of Mechanical and Electronical Engineering, Dezhou University, Dezhou 253023, China\label{addr6}
}

\date{Received: date / Accepted: date}

\abstract{
In this paper, we present the application of a new method measuring Hubble parameter $H(z)$ by using the anisotropy of luminosity distance($d_{L}$) of the gravitational wave(GW) standard sirens of neutron star(NS) binary system. The method has never been put into practice so far due to the lack of the ability of detecting GW. However£¬LIGO's success in detecting GW of black hole(BH) binary system merger announced the potential possibility of this new method. We apply this method to several GW detecting projects, including Advanced LIGO(aLIGO), Einstein Telescope(ET) and DECIGO, and evaluate its constraint ability on cosmological parameters of $H(z)$. It turns out that the $H(z)$ by aLIGO and ET is of bad accuracy, while the $H(z)$ by DECIGO shows a good one. We simulate $H(z)$ data at every 0.1 redshift span using the error information of $H(z)$ by DECIGO, and put the mock data into the forecasting of cosmological parameters. Compared with the previous data and method, we get an obviously tighter constraint on cosmological parameters by mock data, and a concomitantly higher value of Figure of Merit(FoM, the reciprocal of the area enclosed by the $2\sigma$ confidence region). For a 3-year-observation by standard sirens of DECIGO, the FoM value is as high as 170.82. If a 10-year-observation is launched, the FoM could reach 569.42. For comparison, the FoM of 38 actual observed $H(z)$ data(OHD) is 9.3. We also investigate the undulant universe, which shows a comparable improvement on the constraint of cosmological parameters. These improvement indicates that the new method has great potential in further cosmological constraints.
}
\maketitle

\section{Introduction}\label{sect:intro}

In the twenty-first century, we witnessed the bloom of precision cosmology. Precision cosmology even ranked second on a list named "Insights of the decade" from Science magazine in 2010 \cite{2010Sci...330.1615C}. The key of accurate cosmology is to accurately constrain cosmological parameters and their state equations, which can lead us to a better understanding of the evolution of our universe. Four main observations have been developed to constrain cosmological parameters so far \cite{2013PhR...530...87W} : Supernova(SN, \cite{1998AJ....116.1009R}), Baryon Acoustic Oscillation(BAO, \cite{1998ApJ...496..605E}), Galaxy Cluster(CL, \cite{1933AcHPh...6..110Z}), Weak Lensing(WL, \cite{1937ApJ....86..217Z}). Actually, a relatively new tool, observational Hubble parameter data(OHD), is becoming increasingly popular these years because of its effective constraint on cosmological parameters. $H(z)$'s high efficiency lies on the fact that it is the only observation that can directly represent the expanding history of our universe. Compared with the Luminosity distance($d_{L}$) of SN, $H(z)$ contains no integral terms and directly connects with cosmological parameters, which makes it powerful in constraining cosmological parameters, because the integral term can conceal many details and hide important information. As Ma and Zhang \cite{2011ApJ...730...74M} reported, $H(z)$ constrains cosmological parameters much tighter than the same-number SN does. To achieve the same constraint effect of SN subset ConstitutionT, ones need only 64 $H(z)$ data sets under gaussian prior on $H_{0}$,\textbf{ $H_{0}+\sigma_{H_{0}}$}.

There are various ways to detect $H(z)$, which can be generally classified into three types: 1, differential age method \cite{2010JCAP...02..008S}; 2, radial BAO method \cite{2009MNRAS.399.1663G}; 3, standard sirens method \cite{1986Natur.323..310S}. The first two techniques have been employed in the past measurement of $H(z)$, but the number of observed hubble parameter data(OHD) are still insufficient. We get only 38 OHD sets so far, whose accuracies are far from desirable. Now with the development of gravitational wave(GW) detecting technology, it is time to look forwards to the third method: GW standard sirens. GW standard sirens was first proposed and discussed by Schutz \cite{1986Natur.323..310S} . Schutz presented his idea that one can determine the Hubble constant through the observation of GW emitted by decaying orbit of neutron star(NS) binary system. In 2015, although the second generation GW detector Advanced LIGO(aLIGO) operated even not at its design sensitivity, it still detected the first GW signal at its first run \cite{2016PhRvL.116f1102A}.  According to theoretical understanding, GW formula of compact binary system encodes the information of $d_{L}$, providing an access to the direct measurement of $d_{L}$, the crucial parameter we use in this paper. We sense the possibility and potential from detecting gravitational wave.

The detection of GW not only conforms to the general relativity, but also let us see the hope of standard sirens \cite{2017ApJ...848L..13A, 2017PhRvL.119p1101A, 2017Natur.551...85A}. Toshiya Namikawa et al. \cite{2016PhRvL.116l1302N} studied GW standard sirens as a cosmological probe without redshift. But it is hard to get the corresponding redshift information without electromagnetic counterpart. The absence of redshift impedes some further research. If one were given the $H(z)$ and its corresponding redshift, the scope of research would be wildly broaden. In 2006, a new way to narrow the relative error of $H(z)$ by the dipole of $d_{L}$ has been proposed \cite{2006PhRvL..96s1302B}. The relative error is measured by the dipole of SN. But the problem is that the new method needs plenty of SNs if we want to get a relatively accurate $H(z)$, which can not be met in reality. Although this method also has problem in measuring high-$z$ $H(z)$, it is an instructive idea. Atsushi Nishizawa and Atsushi Tamga \emph{et al.}\cite{2011PhRvD..83h4045N} gave us an alternative by pointing it out that we can get information of $d_{L}$ through the gravitational wave function of NS binary system, instead of SN . We follow his idea and choose NS binary system as our research subject in this paper, because a rough estimate would tell us that the number of observed NS binary system turns out much bigger than that of SN. NS binary system can help us dramatically narrow down the statistical error.

Pozzo \cite{2012PhRvD..86d3011D} proposed a general Bayesian theoretical framework for cosmological inference, which can conveniently include the prior information about the GW source. This framework defines the likelihood based on the difference between the strain of each detector and the GW template, and the posterior probability distribution for the cosmological parameters is calculated through the quasi-likelihood obtained by marginalizing over the GW signal intrinsic parameters. Applying the framework the author constrained the Hubble constant $H_0$ to an accuracy of $~4-5\%$ at $95\%$ confidence. Nearly all subsequent work of using GW sources for cosmological inference is based on this Bayesian framework. The same framework was adopted by Taylor et al. \cite{2012PhRvD..85b3535T}, but the likelihood was defined on the assumption that the number count of GW events detected by a detector is a Poisson distributed random variate. They measured the Hubble constant using GW signals of NS binaries by narrowing the distribution of masses of the underlying NS population. That is, $H_0$ was determined to $\pm10\%$ using $\sim 100$ observations. By assuming that the masses of NS binaries can be modeled by a Gaussian distribution and that both masses of the double NS systems are equal, the authors found their chirp masses are approximately normally distributed and got the corresponding mean and standard deviation. Then, using the same method, they explored the prospects for constraining cosmology using GW observations of neutron star binaries by the proposed Einstein Telescope (ET), a third-generation ground-based interferometer. This time they fixed $H_0$, $\Omega_{m,0}$ and $\Omega_{\Lambda,0}$ and constrained the dark-energy equation of state (EOS) parameters \cite{2012PhRvD..86b3502T}. With a $10^5$-event catalog, they constrained the dark-energy EOS parameters to an accuracy similar to forecasted constraints from future CMB+BAO+SNIa measurements. Chen et al. \cite{2018Natur.562..545C} investigated the measurement of Hubble constant at various cases: with and without electromagnetic counterpart, binary NS mergers and binary black hole mergers. They showed that that LIGO and Virgo can be expected to constrain the Hubble constant to a precision of $~2\%$ within 5 years and $~1\%$ within a decade. Vitale and Chen \cite{2018PhRvL.121b1303V} dealt with neutron star black hole mergers and focused on measuring the luminosity distance to a source. They concluded that the $1-\sigma$ statistical uncertainty of the luminosity distance for spinning black hole neutron star binaries can be up to a factor of $\sim10$ better than for a non-spinning binary neutron star merger with the same signal-to-noise ratio. Pozzo et al. \cite{2017PhRvD..95d3502D} investigated the accuracy of the measured cosmological parameters using information coming only from the gravitational wave observations of binary neutron star systems by the Einstein Telescope. They used Fisher matrix method to extract redshift information of a source given that information about the equation of state of the source is available \cite{2012PhRvL.108i1101M}. They found by direct simulation of $10^3$ detections of binary neutron stars, $H_0$,
$\Omega_m$, $\Omega_\Lambda$, $w_0$ and $w_1$ can be measured at the $95\%$ level with an accuracy of $\sim8\%$, $65\%$, $39\%$, $80\%$ and
$90\%$, respectively. Different to the previous studies that focussed on constraining the parameters of specific cosmological models, our work emphasises a model independent measurement of $H(z)$. A model free approach will generally produce a weaker constraint on any particular model than the model-specific analysis, but it has more flexibility if the true model deviates from the model assumed.

For the current observational status of GW, several frequency windows of its are targeted by different detectors. The second generation detector are mainly aimed at frequency window $10 \sim 1000$Hz, such as aLIGO and VIRGO. The next generation detector plan to reach lower frequency band. The project DECIGO was designed most sensitive at $0.1 \sim 10$Hz, while the Einstein Telescope(ET) may also reach the frequency $\sim 1$Hz. The space-based eLISA can even detect GW of $10^{-4} \sim 10^{-1}$Hz. In this paper, we make use of GW sirens to measure $H(z)$ by estimating the error of $d_{L}$, a little different from the method proposed by Schutz \cite{1986Natur.323..310S}. Because NS binary system are used as the source of GW in this paper, the GW signal frequency of whom mostly ranges in $10 \sim 1000$Hz, we ignore the projects whose optimal sensitivity are far away from $10 \sim 1000$Hz, such as eLISA, and choose the ones whose optimal sensitivity locate around $10 \sim 1000$Hz. Finally, aLIGO, Einstein Telescope(ET) and DECIGO are chosen as our research subject.

Most importantly, the Advanced LIGO and Advanced Virgo gravitational-wave detectors made their first observation of a binary neutron star inspiral, and detected the signal of GW170817 with a combined signal-to-noise ratio of 32.4\cite{2017PhRvL.119p1101A,2017Natur.551...85A}. In addition it provides the first direct evidence of a link between binary neutron star mergers and short $\gamma$-ray bursts. The combined analyses of the gravitational-wave data and electromagnetic emissions are providing new insights into independent tests of cosmological models, so GW170817 marks the beginning of a new era of cosmology. Using the data of GW170817, they performed the gravitational-wave standard siren measurement of the Hubble constant \cite{2017Natur.551...85A} to be $70^{+12}_{-8}\rm{kms^{-1}Mpc^{-1}}$. Different from their works, in this paper, we focus mainly on two aspect: 1, How will it work out if we apply the new method to some projects? 2, how about the quality of the $H(z)$ by this method, or to what degree could we constrain cosmological parameters? This paper is organized as follows. In sec 2, we sketch the idea of GW standard sirens method, and apply it to some GW detecting projects. In sec 3, we simulate the $H(z)$ data, and analyze the constraint ability of the mock data for $\Lambda$CDM and the undulant universe. In sec 4, we discuss the result and talk a little about the corresponding redshift. All through this paper, we adopt the natural unit, $c=G=1$.


\section{Method}\label{sect:data}

\subsection{Dipole of luminosity distance}

\begin{figure}[tbp]
\centering
    \includegraphics[width=0.5\textwidth,bb=0 0 560 420]  {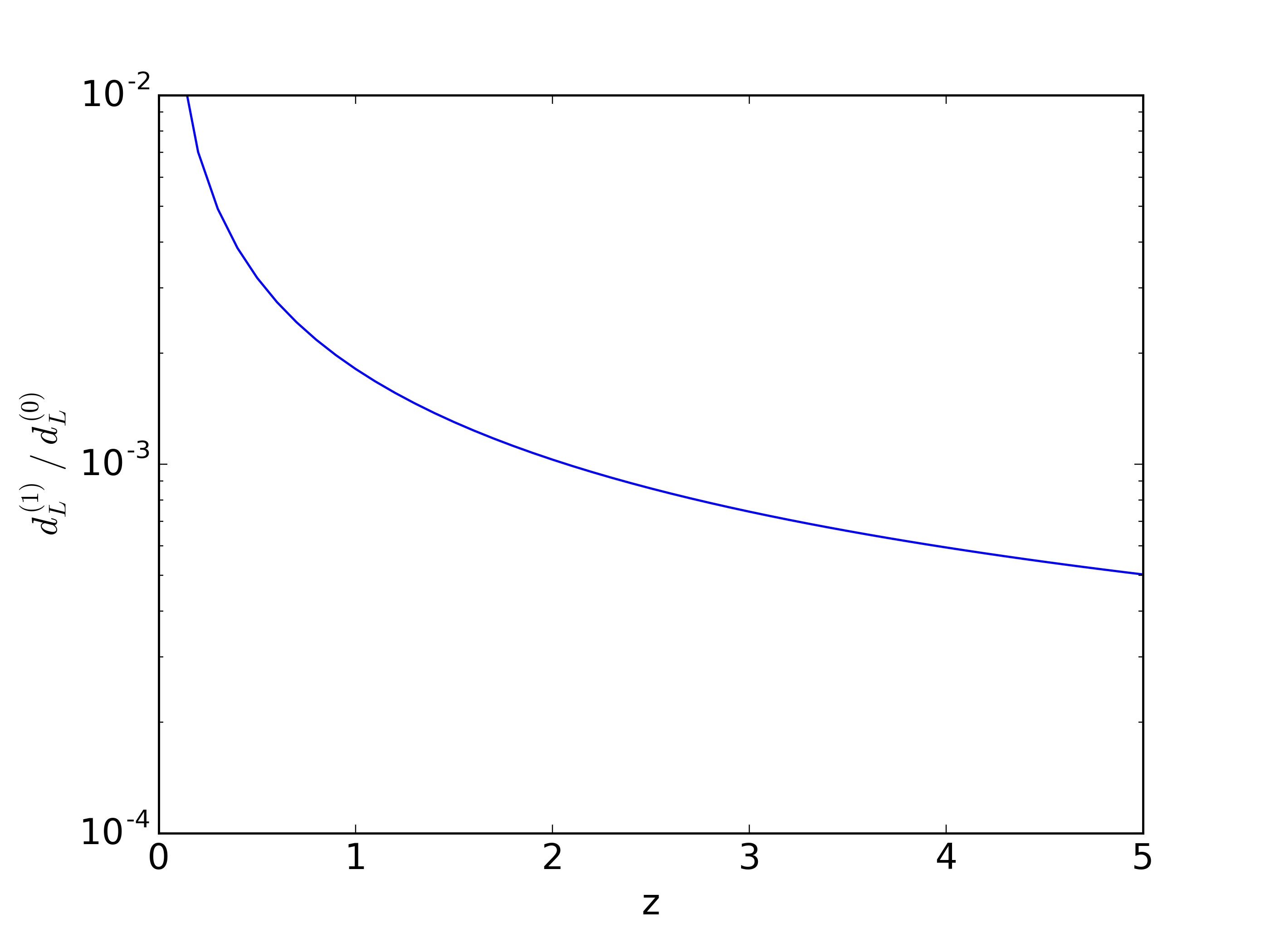}
    \caption{ The value of $d_{L}^{(1)}/d_{L}^{(0)}$ at different redshift for $v_{0}=369 \rm{km/s}$ \cite{2011ApJS..192...14J}. As shown in the picture, the ratio goes very large , even bloom up, at low redshift. That is caused by the fact that the ratio approximate to $(1+z)|v_{0}|/z$ at the limit of $z=0$.}
    \label{1}
\end{figure}

If the universe is completely homogeneous and isotropic on large scale, and the observer is relatively rest with the cosmic microwave background(CMB), the luminosity distance, $d_{L}$, would be just the same form and has the same expression as in standard cosmology. But in fact, there are perturbations around ideal condition leading into the appearance of correction term of $d_{L}$ \cite{1987MNRAS.228..653S}. Therefore $d_{L}$ can be written as follow:
\begin{equation}
    d_{L}=d_{L}^{(0)}+d_{L}^{(1)}+ \text{higher order terms,}
\end{equation}
where $d_{L}^{(0)}$ represents the traditional meaning of luminosity distance in unperturbed Friedmann universe, also the average of $d_{L}$ on all direction, and $d_{L}^{(1)}$ means the dipole of $d_{L}$. The contribution to higher order terms coming from the weak gravitational lensing effect is so small compared with dipole that we ignore them here \cite{2006PhRvD..73b3523B}. The dipole is dominated by the peculiar velocity of observers. If you want to check it or feel intrigued by the theory, you can look up the reference for the details \cite{2006PhRvL..96s1302B}. Here is the final result:

\begin{equation}
 d_{L}^{(1)}=\dfrac{(1+z)^{2}}{H(z)} |v_{0}|;
 \dfrac{\Delta H(z)}{H(z)}=\sqrt{3} \left[ \dfrac{d_{L}^{(1)}}{d_{L}^{(0)}} \right]^{-1} \left[ \dfrac{\Delta d_{L}^{(0)}}{d_{L}^{(0)}} \right]   \text{,}
\end{equation}
where $|v_{0}|$, $z$, $H(z)$, \textbf{$\Delta H(z)$} respectively denote the projection of observer peculiar velocity on the direction of sight, the redshift of the observed celestial body, the expanding rate at the redshift $z$, the absolute error of $H(z)$, and $\Delta d_{L}^{(0)}$, $\Delta d_{L}^{(1)}$ means the error of $d_{L}^{(0)}$, $d_{L}^{(1)}$ respectively. From the equations above,given the value of $d_{L}^{(1)}/ d_{L}^{(0)}$ and $\Delta d_{L}^{(0)}/d_{L}^{(0)}$, $\Delta H(z)/H(z)$ can be easily calculated. The result of the term $d_{L}^{(1)}/d_{L}^{(0)}$ is shown in Fig. \ref{1}. To get $\Delta H(z)/H(z)$, the only remaining problem is to find out $\Delta d_{L}^{(0)}/d_{L}^{(0)}$, which can be solved by analyzing observed GW function in following subsection. Also, the mean error $\Delta H(z)$ reduces to $\Delta H(z)/\sqrt{N}$ if we observe N independent sources at the given redshift. Thus, we can improve the accuracy by the observation of a large number of sources.

\subsection{GW standard sirens}

One can use SN to illustrate the method of reducing the error of $H(z)$. But due to the number and distribution of SN, it doesn't work well, especially at high-z region. Considering the advantage of the larger number of observed sources, which can dramatically narrow down the error of $H(z)$, we choose NS binary system as an alternative of SN. There is another problem for black hole binary system: black hole seldom radiates electromagnetic wave, rendering it hard to measure its corresponding redshift up to now. This is an important factor to choose NS binary system.

In GW experiments, one can extract the property of the source and cosmological information by comparing detected waveform with theoretical template.  That is  what LIGO team did when the first detected the GW of two back holes merged \cite{2016PhRvL.116f1102A}. The typical Fourier transform of GW waveform can be expressed by
\begin{equation}
    \widetilde{h}(f)=\dfrac{A}{d_{L}(z)}M_{z}^{5/6}f^{-7/6}e^{i\Psi(f)}\text{,}
\end{equation}
which is based on the average over sky location. Here $A=(\sqrt{6}\pi^{2/3})^{-1}$ is a constant geometrically averaged over the inclination angle of a binary system. $d_{L}(z)$ is the luminosity distance at redshift $z$, and we can set it as $d_{L}^{(0)}$ because we need to observe plenty of source at the given redshift. $M_{z}=(1+z)\eta^{3/5}M_{t}$ with the definition of total mass $M_{t}=m_{1}+m_{2}$ and symmetric mass ratio $\eta=m_{1}m_{2}/M_{t}^{2}$. The last unknown function $\Psi(f)$ is a little intricate. It is the frequency-dependent phase caused by orbital evolution. Usually we deal with it by post-Newtanion(PN) approximation, an approximation to general relativity in the weak-field, slow-motion regime \cite{1993PhRvD..47.3281K}. Its concrete expression will not affect the final result, because this term will be eliminated when we do the following calculation. Here we just need to know that it is a function of the coalescence time $t_{c}$, the phase when emitted $\phi_{c}$, $M_{z}$, $f$, $\eta$.

There are five unknown parameters, namely: $M_{z}$, $\eta$, $t_{c}$, $\phi_{c}$, $d_{L}$, where $d_{L}$ is the only parameter that has nothing do with the own property of binary system. For the convenience of calculating, we just take account of equal mass NS binary system with $1.4M_{\bigodot}$, and set $t_{c}=0$, $\phi_{c}=0$. Then we have $M_{z}=1.22(1+z)M_{\bigodot}$, $\eta=0.25$. Though GW may tell us some information about the redshift \cite{2012PhRvL.108i1101M, 2014PhRvX...4d1004M}, we have no data about the redshift and we need a general method to get the reshift information. We should still resort to electromagnetic observation to find out corresponding redshift. Cutler and Holz \cite{2009PhRvD..80j4009C} demonstrated its technological viability.

\begin{figure}[htb]
\centering
    \includegraphics[width=0.5\textwidth,bb=0 0 560 420]  {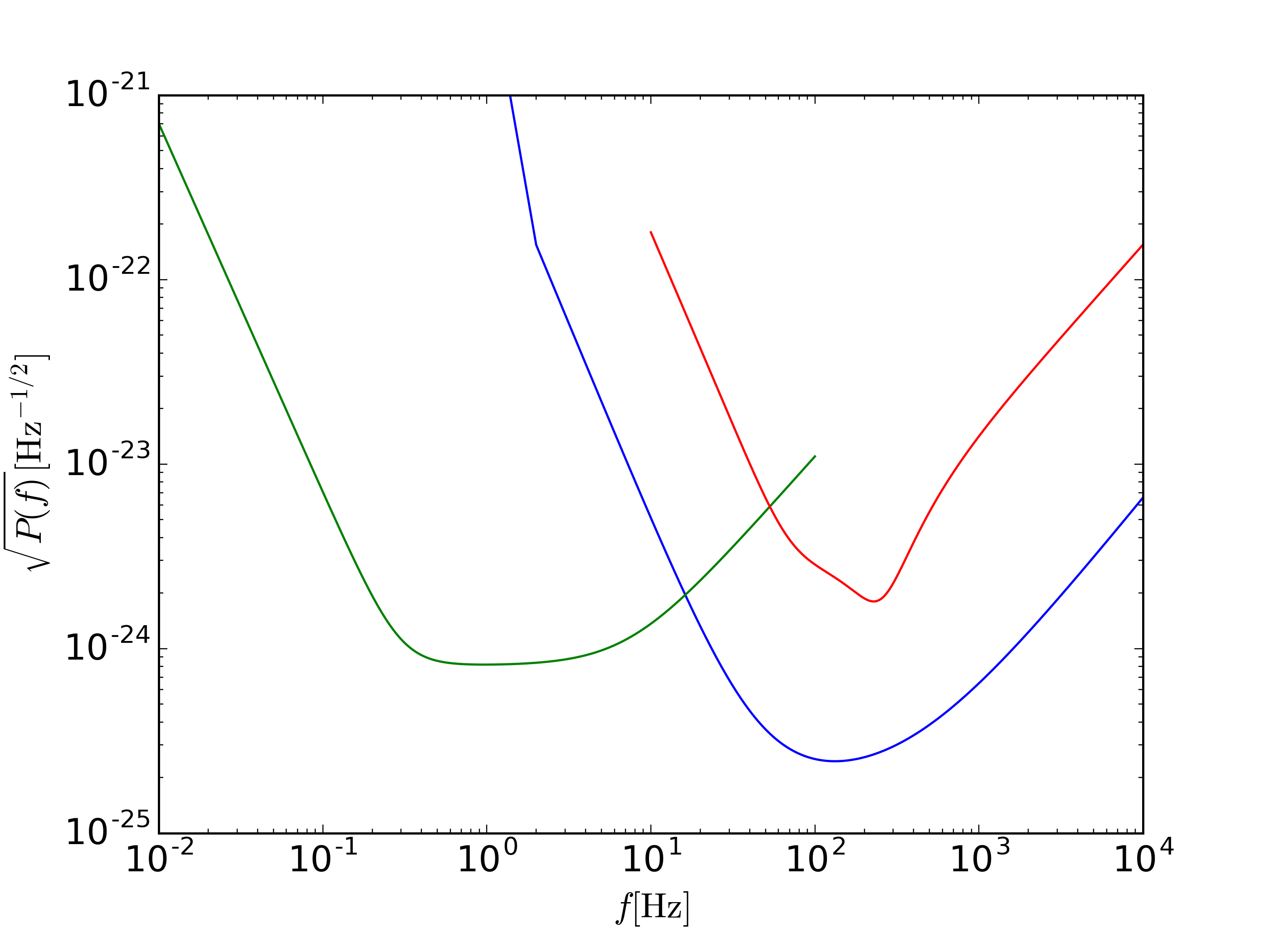}
    \caption{\label{2} The noise power spectrum. Green curve represents $P_{1}(f)$ for DECIGO, blue curve represents $P_{2}(f)$ for ET, red curve represents $P_{3}(f)$ for aLIGO respectively.}
\end{figure}

We use fisher matrix to estimate error. Fisher matrix has its limit: the Cramer-Rao bound, and it's break down at low SNR. The error estimate for $d_{L}$ is based on Fisher matrix that is given by
\begin{equation}
 \Gamma_{ab}=4 Re \int_{f_{min}}^{f_{max}}  \dfrac{ \partial_{a} \tilde{h}^{*}_{i}(f)   \partial_{b} \tilde{h}_{i}(f)    }{P(f)}  df \text{,}
\end{equation}
where $\partial_{a}$ means derivative with respect to parameter $\theta_{a}$. For DECIGO, there exist eight interferometric signals, $\Gamma_{ab}$ should multiplied by 8. We set values to parameters expect for $d_{L}$, so the only parameter in $\Gamma_{ab}$ is  $d_{L}$. $P(f)$ is the noise power spectrum, and the $P(f)$ for DECIGO, ET and aLIGO are shown in Fig. \ref{2}. Here we give the expression of each detector's noise curve, $P_{1}(f)$, $P_{2}(f)$, $P_{3}(f)$ for DECIGO, ET and aLIGO respectively.\\

\textbf{DECIGO}: DECIGO(Deci-Hertz Interferometer Gravitational Wave Observatory) is a planed space-based GW observation aimed at $0.1\sim10$Hz frequency window. Its configuration is still to be decided. Here we adopt the following parameters in its configuration: the arm length $1000$km, the output laser power $10$W with wavelength $\lambda=532$nm, the mirror diameter 1m with its mass 100Kg, and the finesse of FP cavity 10. Thus its noise curve is \cite{2006CQGra..23S.125K}
\begin{equation}
\begin{split}
P_{1}(f)=&6.53\times10^{-49} \left[1+(\dfrac{f}{7.36Hz})^{2} \right] \\
&+4.45\times10^{-51}\times (\dfrac{f}{1Hz})^{-4}\times \dfrac{1}{1+(\dfrac{f}{7.36Hz})^{2}}  \\
&+4.94\times10^{-52}\times (\dfrac{f}{1Hz})^{-4}  \rm{Hz^{-1}} \text{.}
\end{split}
\end{equation}

\textbf{ET}: ET(Einstein Telescope) is a third generation GW detector, whose design is not finished. Here we just consider the simplest case with 10km arms.  We adopt the fitting expression given by Keppel and Ajith \cite{2010PhRvD..82l2001K}
\begin{equation}
\begin{split}
P_{2}(f)= &1\times10^{-50} [2.39\times10^{-27}(\dfrac{f}{100Hz})^{-15.64}   +0.349\times(\dfrac{f}{100Hz})^{-2.145}  \\ &+1.76\times(\dfrac{f}{100Hz})^{-0.12}+0.409\times(\dfrac{f}{100Hz})^{1.1}    ]^{2}   \rm{Hz^{-1}}  \text{.}
\end{split}
\end{equation}

\begin{figure}[tbp]
    \includegraphics[width=0.5\textwidth,bb=0 0 560 420]  {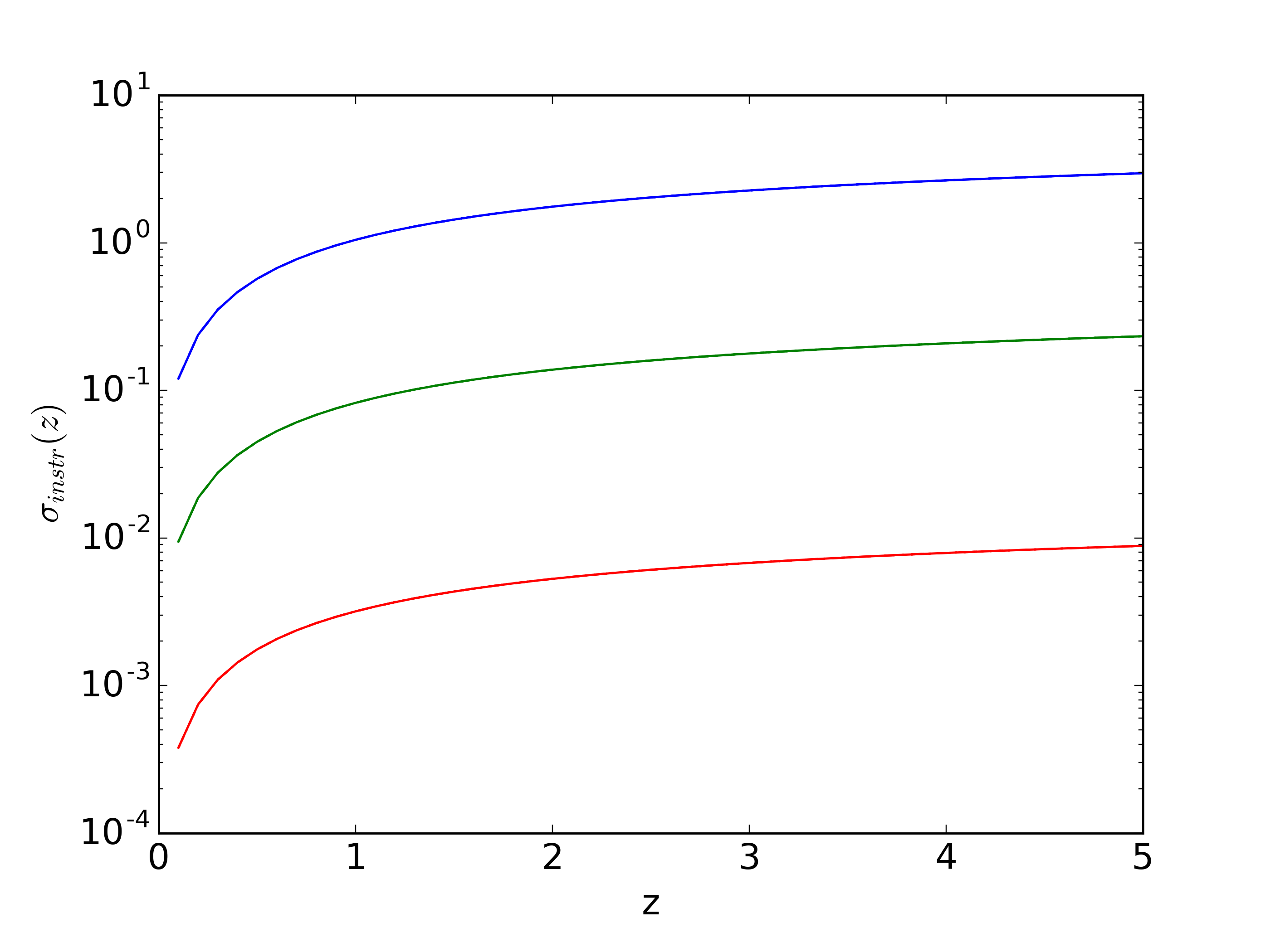}
    \caption{\label{3} The device error $\sigma_{instr}(z)$ for DECIGO(Red line), ET(green line) and aLIGO(blue line)
    respectively.}
\end{figure}

\textbf{aLIGO}: aLIGO is an available second generation detector whose optimal sensitivity band match with the frequency window of GW from NS binary system . The first run of aLIGO did not reach its design sensitivity. Here we use the noise curve fitted by \cite{2005PhRvD..71h4008A}. It is not an accurate expression, but an approximation of the original curve is given by \cite{2002gr.qc.....4090C}
\begin{equation}
\begin{split}
P_{3}(f)= &1\times 10^{-49} [ (\dfrac{f}{215Hz})^{-4.14}-5\times(\dfrac{f}{215Hz})^{-2}  \\
                    &+111\times( \dfrac{1-(\dfrac{f}{215Hz})^{2}+(\dfrac{f}{215Hz})^{4} /2 }{  1+(\dfrac{f}{215Hz})^{2}/2}   )  ]  \rm{ Hz^{-1}} \text{.}
\end{split}
\end{equation}
In the expression of $\Gamma_{ab}$, the lower cutoff of frequency, $f_{min}$, is a function of observation time $T_{obs}$,
$f_{min}=0.233( 1M_{\bigodot} / M_{z} )^{5/8}( 1yr / T_{obs} )^{3/8}Hz$. In the case of our paper, for a given $T_{obs}$, $f_{min}$ changes little with $M_{z}$, which is always in the high strain noise region. It makes no big difference to the result of the integral. A reasonable setting of the value of $f_{min}$ will work. But for prudence, we just take the original expression of $f_{min}$ when calculating the integral. And the higher cutoff, $f_{max}$, is the inner-most stable circular orbit frequency, whose typical value is of ~kHz order \cite{2013CQGra..30l3001B}. More specific, $f_{max}\simeq 2000$Hz in our case. In the calculation, $f_{max}$ can be set by the property of the integrand. The value of integrand sharply drops with $f$ getting larger, so its contribution to $\Gamma_{ab}$ can be neglected. For the reason of integrand property, we set the $f_{max}$ of DECIGO, ET, aLIGO respectively to be 100Hz, 2000Hz and 2000Hz. Then the one-sigma instrument error is
\begin{equation}
    \sigma^{\theta_a}_{instr}(z)=\Delta\theta_{a}=\sqrt{\{\Gamma^{-1}\}_{aa}} \text{.}
\end{equation}
If we launch an observation for a given source, the one-sigma error estimate $\sigma_{instr}$ of $d_{L}$ arises from instrumental noise. For a given device, no matter it is DECIGO, ET or aLIGO, the accuracy of $d_{L}$ is the same even for different observation time. It makes no difference for the $\sigma_{instr}$ no matter how long the observation continues, which is mainly because that the error is due to the property of device. The observation is band-limited. The source is visible only for the time it takes to move form the low frequency list of the detector's sensitivity to merger. For any observation longer than that time the precision is the same since you do not observe the source any more. The $\sigma_{instr}$ of three devices are showed in Fig. \ref{3}. As we can see, the accuracy is far from desirable. The $H(z)$ error would increase if we include other errors . We need to take measure to narrow down the error. This is what we do in next subsection

\subsection{H(z) error}

In last subsection, we already calculate the device error $\sigma_{instr}$ for a given NS binary system under a given observation device. Besides the device error-the dominating error, there are two main errors, namely the lensing error and the peculiar-velocity error. The lensing error is due to the lens effect. Here we take a recent fitting by \cite{2010PhRvD..81l4046H},
\begin{equation}
   \sigma_{lens}(z)=0.066*[\frac{1-(1+z)^{-0.25}}{0.25}]^{1.8}.
\end{equation}
 And the peculiar-velocity error is a kind of Doppler effect of the movement of the celestial body, essentially. The peculiar-velocity error can be described as (\cite{2007PhRvL..99h1301G})
\begin{equation}
    \sigma_{pv}(z)=| 1- \frac{(1+z)^2}{H(z)d_{L}(z)}|\sigma_{v,gal},
\end{equation}
where $\sigma_{v,gal}= 300$km/s is the approximation of the 1-dimensional velocity dispersion of the galaxy. Then we get the expression of relative error of $d_{L}(z)$:

\begin{equation}
  [\dfrac{\Delta d_{L}^{(0)}}{d_{L}^{(0)}}]^{2}= \sigma_{instr}^{2}(z)+\sigma_{lens}^{2}(z)+\sigma_{pv}^{2}(z).
\end{equation}

Before we do the calculation to get the relative error of $H(z)$, there is one more step we can do for a better accuracy. The mean error will statistically abate if we have many independent sources. Reducing the error of $H(z)$ by observing many NS binary system at the same redshift may be feasible. The problem is to what degree can we reduce the error? First we need to figure out the number distribution $\dot{n}(z)$ of NS binary system at different redshift. The distribution of NS binary system can be described and estimated. According to Cutler and Harms \cite{2006PhRvD..73d2001C}, the fitting of NS-NS merger rate can be given by:
\begin{equation}
    \dot{n}(z)=\dot{n}_{0}s(z), s(z)=\begin{cases} 1+2z, &z\leq1 \\ 0.75(5-z), &1<z<5 \\ 0, &z\geq5  \text{,} \end{cases}
\end{equation}
where s(z) is estimated from star formation history inferred from UV luminosity, and $\dot{n}_{0}$ represents the merger rate at present time. Then $\Delta N$, the number of NS-NS merger at redshift bin $\Delta z$, is expressed by:
     $\Delta N(z)=T_{obs}  \int_{z-\Delta z/2}^{z+\Delta z/2}  4\pi$ \\
     $[d_{L}(z')/(1+z')]^{2} \dot{n}(z')/(1+z') /H(z')    dz' $.

\begin{figure}[tbp]
    \includegraphics[width=0.5\textwidth,bb=0 0 560 420]{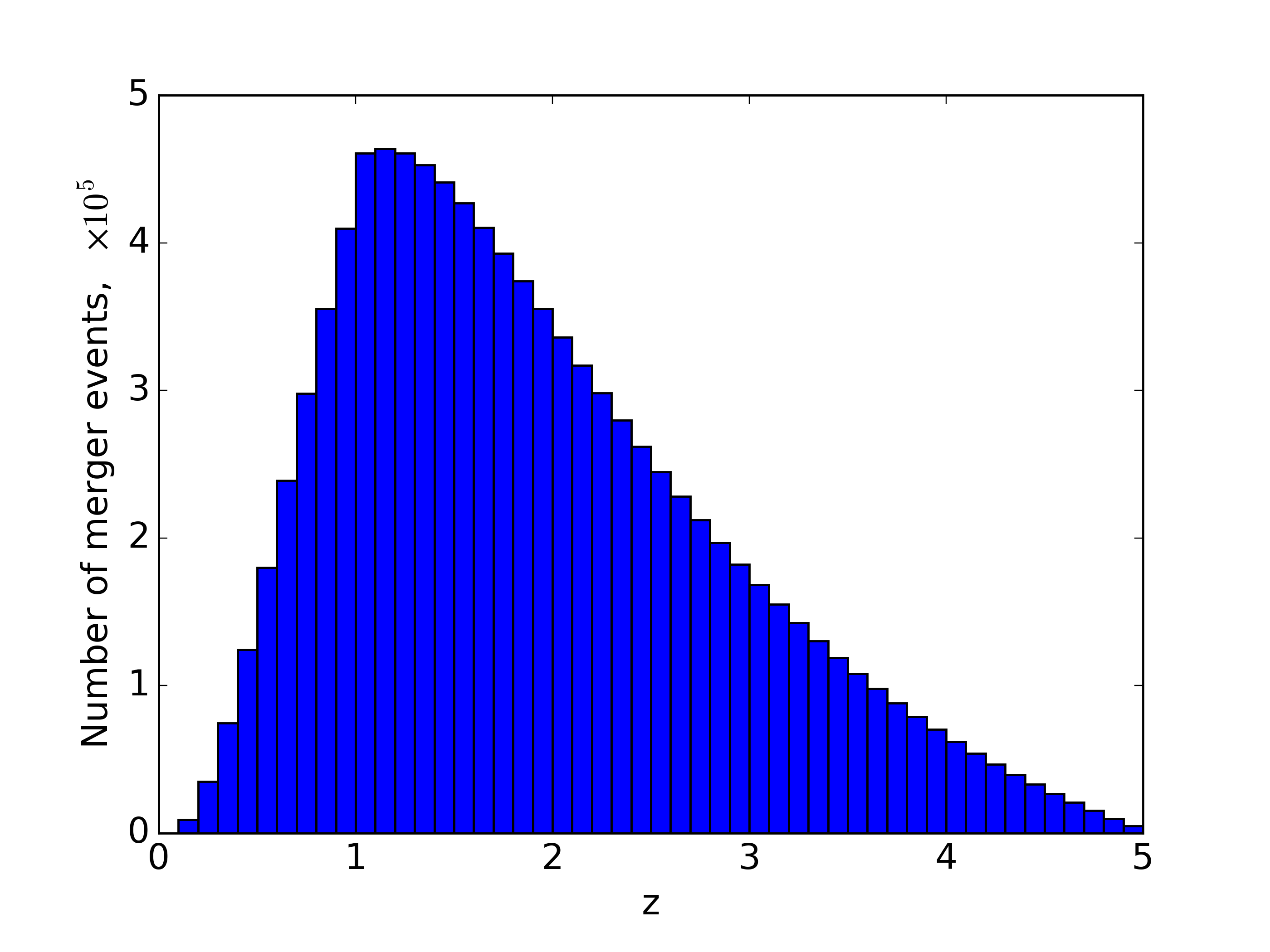}
    \caption{\label{4} The number of merger events in each redsshift bin of $\Delta z$=0.1 at a redshift $z$ during 10-year observation.}
\end{figure}

Recent work doesn't provide solid evidence of the exact value of $\dot{n}_{0}$. he latest $\dot{n}_{0}$ range inferred by the observation of GW170817 is $0.32-4.7 \times 10^{-6}\rm{Mpc^{-3}yr^{-1}}$ \cite{2017PhRvL.119p1101A}. Also not every merger event would be detected. Here we encounter an conundrum. Considering that we are aimed at evaluating the method, not launching an actual observation here, we decide to, a bit arbitrarily, set $\dot{n}_{0}$ equal to $1.0 \times 10^{-6}\rm{Mpc^{-3}yr^{-1}}$, and assume all the merger events could be detected, and the redshift width $\Delta z=0.1$. Thus we get the estimate of 10-year observed number of NS binary system merger at different redshift, which is shown in Fig. \ref{4}.
\begin{figure}[htb]
    \includegraphics[width=0.5\textwidth,bb=0 0 420 280]  {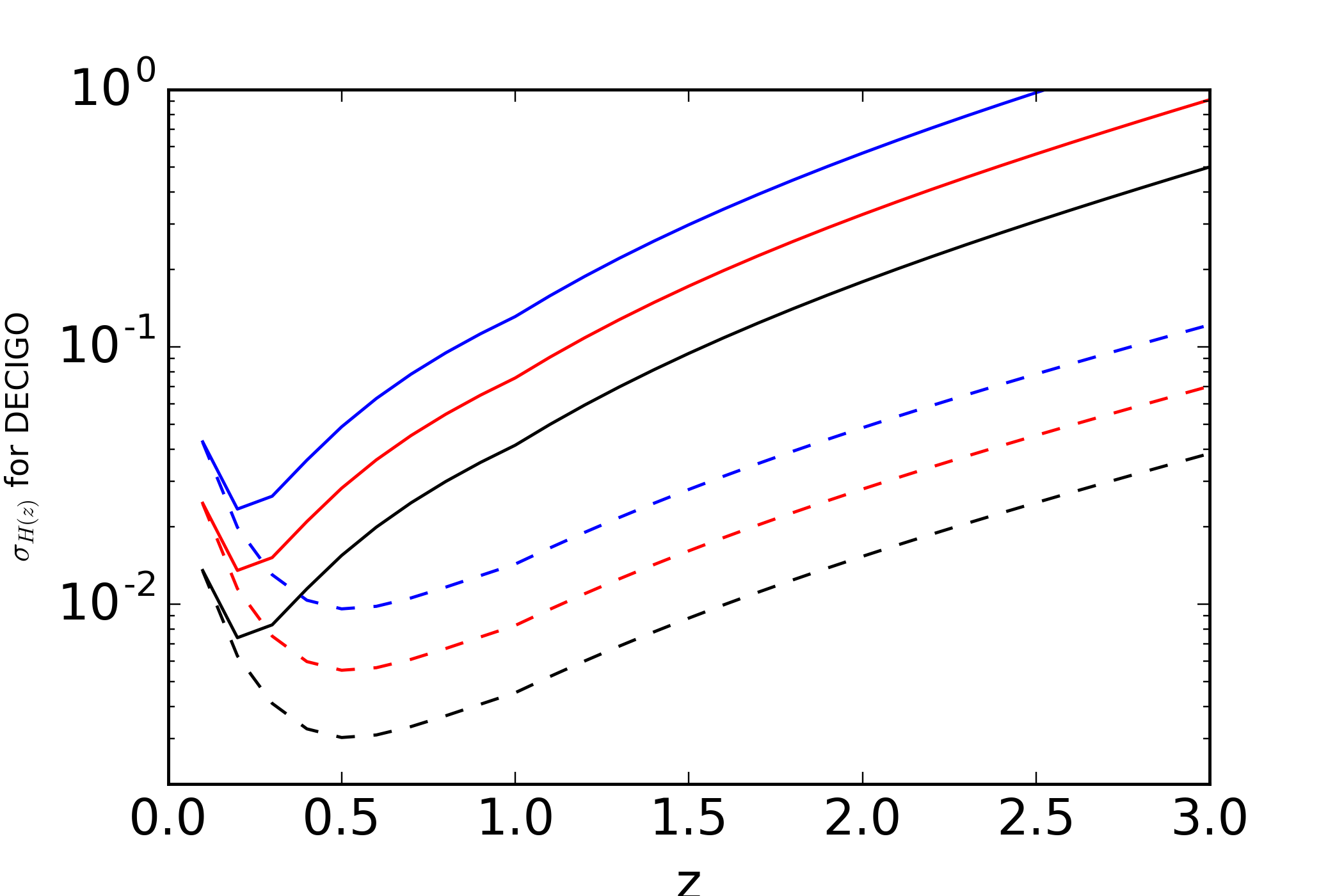}
    \caption{\label{5}  Relative error of $H(z)$ for DECIGO. The dash lines and the solid lines indicate relative errors without and with lens errors respectively. The different colors of of the lines represent different observation time, blue for 1-year, red for 3-year, black for 10-year observation respectively}
\end{figure}
\begin{figure}[htb]
    \includegraphics[width=0.5\textwidth,bb=0 0 420 280]  {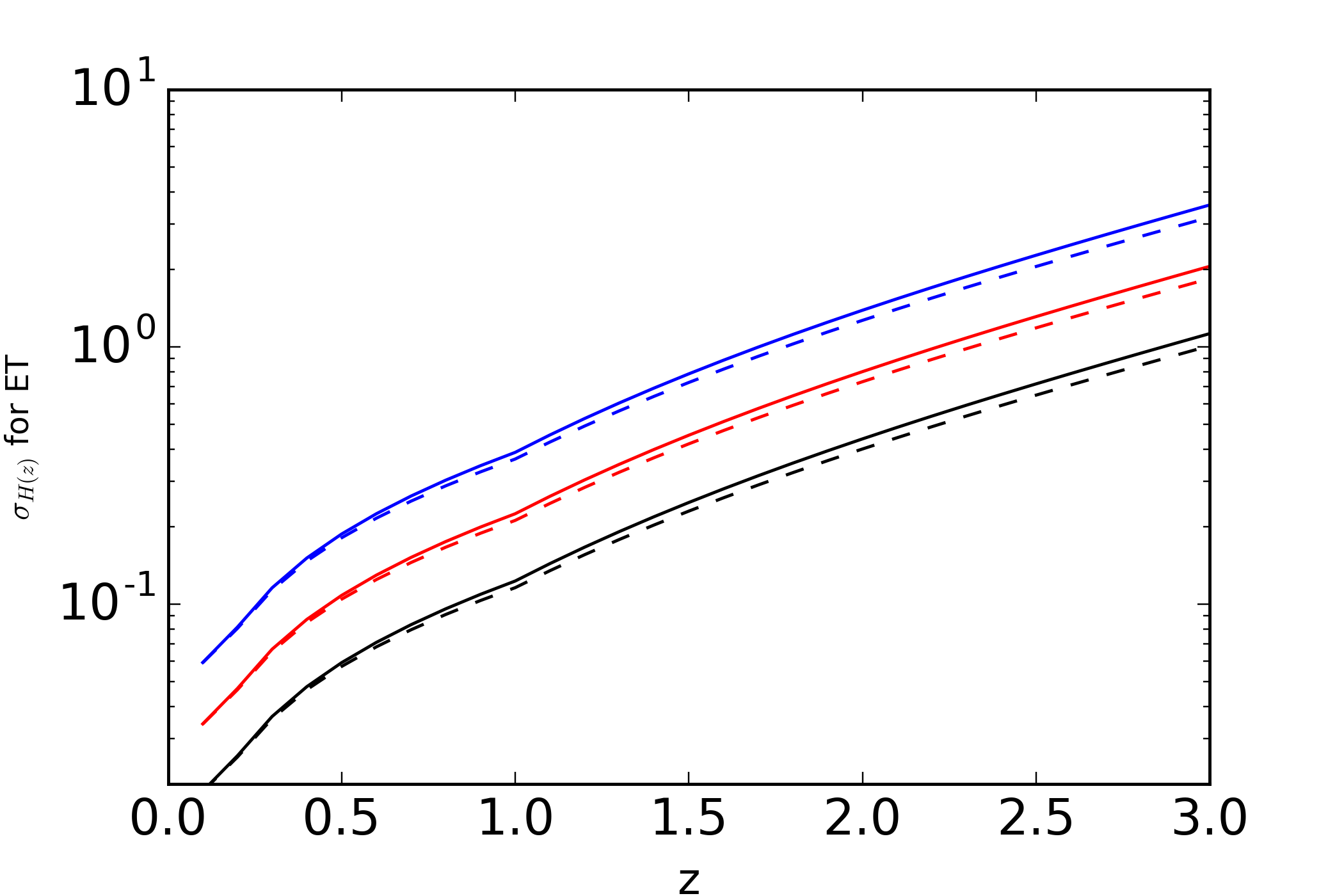}
    \caption{\label{6} Same as Fig. \ref{5}, but for ET}
\end{figure}

\begin{figure}[htb]
    \includegraphics[width=0.5\textwidth,bb=0 0 420 280]  {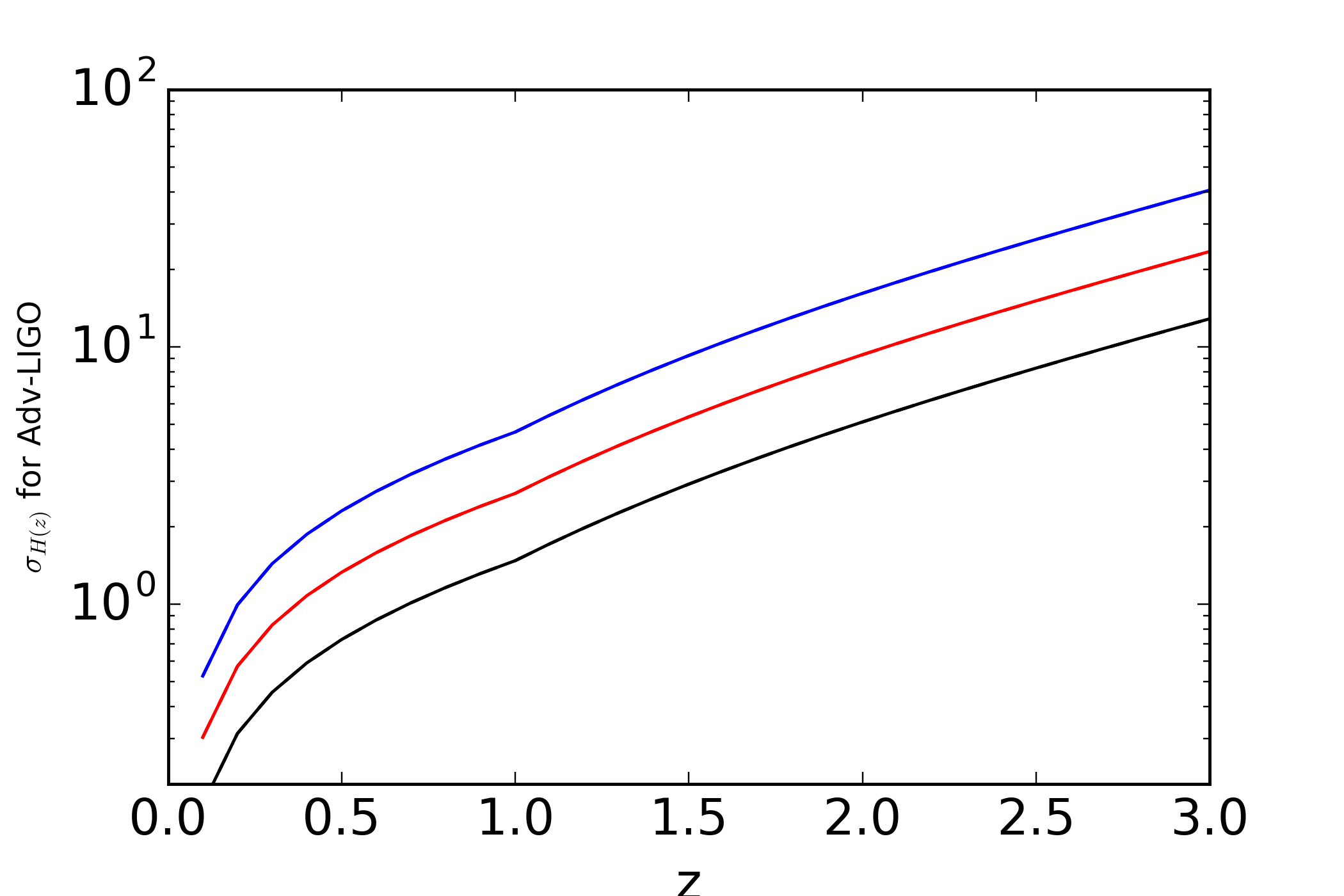}
    \caption{\label{7} Same as Fig. \ref{5}, but for aLIGO. For a given observation time, the dash line and the solid line overlaps, because the lens error is relatively small compared with the instrument error of aLIGO}
\end{figure}

The total number of SN is just of hundred-magnitude by now, while the observed number of NS-NS merger event will be much larger than that of SN, showing a tremendous potential in reducing the mean error of $H(z)$. And from above equation, the number of NS-NS merger at fixed redshift increases with $T_{obs}^{1/2}$. The elongation of observation time can remedy the drawback of the device sensibility.

Combining with the information we get above, we can calculate the $H(z)$ error for a specific device under a given observation time now. The relative error of $H(z)$ by DECIGO, ET and aLIGO is shown in Fig. \ref{5}, Fig. \ref{6}, and Fig. \ref{7}, for 1-year, 3-year, 10-year observation respectively. The relative error by aLIGO is a total disaster, which basically has little application value in constraining cosmological parameters. The error by ET is a little better, especially at low redshift region, because ET is more sensitive than aLIGO. Thus DECIGO plays best in this method. When redshift reach 3, due to the decreasing of the number of observed NS-NS merger event with redshift, the relative error of $H(z)$ is magnified, but still quite small. And the elongation of $T_{obs}^{1/2}$ shows a great ability in narrowing down the error. We stress here that $\sigma_{lens}(z)$ contributes a lot to the distance error $\Delta d_{L}^{(0)}/d_{L}^{(0)}$. Of course, various techniques have been developed to reduce $\sigma_{lens}(z)$. Stefan Hilbert \cite{2011MNRAS.412.1023H} suggested that deep shear survey can narrow the lens error. Hirata \cite{2010PhRvD..81l4046H} found that ones can improve the distance determination typically by a factor of 2-3 by exploiting the non-Gaussian nature of the lens magnification distribution. C. Shapiro \cite{2010MNRAS.404..858S} used the procedure 'delensing', to estimate the magnification and thereby remove it by a weak lensing map. It may be too optimistic to remove all the lens error. But we can rely on it that we could reduce the lens error to a very low level in the near future. Therefore, for simplicity, we will ignore $\sigma_{lens}(z)$ in the following sections.


\section{Evaluation}

\subsection{Simulating data}
\label{sec:morph}

The new method for measuring $H(z)$ has been presented and its error analysis has been done above. The problem is how $H(z)$ can accurately observed by this way to constrain cosmological parameters? So far, we did not obtain actual OHD by this way. But it doesn't necessarily mean we can do nothing about it. A reasonable and rational simulation can help us forecast and evaluate. Since $H(z)$ by aLIGO has a bad accuracy, we carry on no simulation and forecast for aLIGO here. $H(z)$ by ET can do some simulation and forecast. The problem is that the effect is a little bad, even worse than 38 OHD sets. We do not plan to show it here, too. Thus, DECIGO is the only device we discuss in following sections.

\begin{figure}[htb]
    \includegraphics[width=0.5\textwidth,bb=0 0 560 420]  {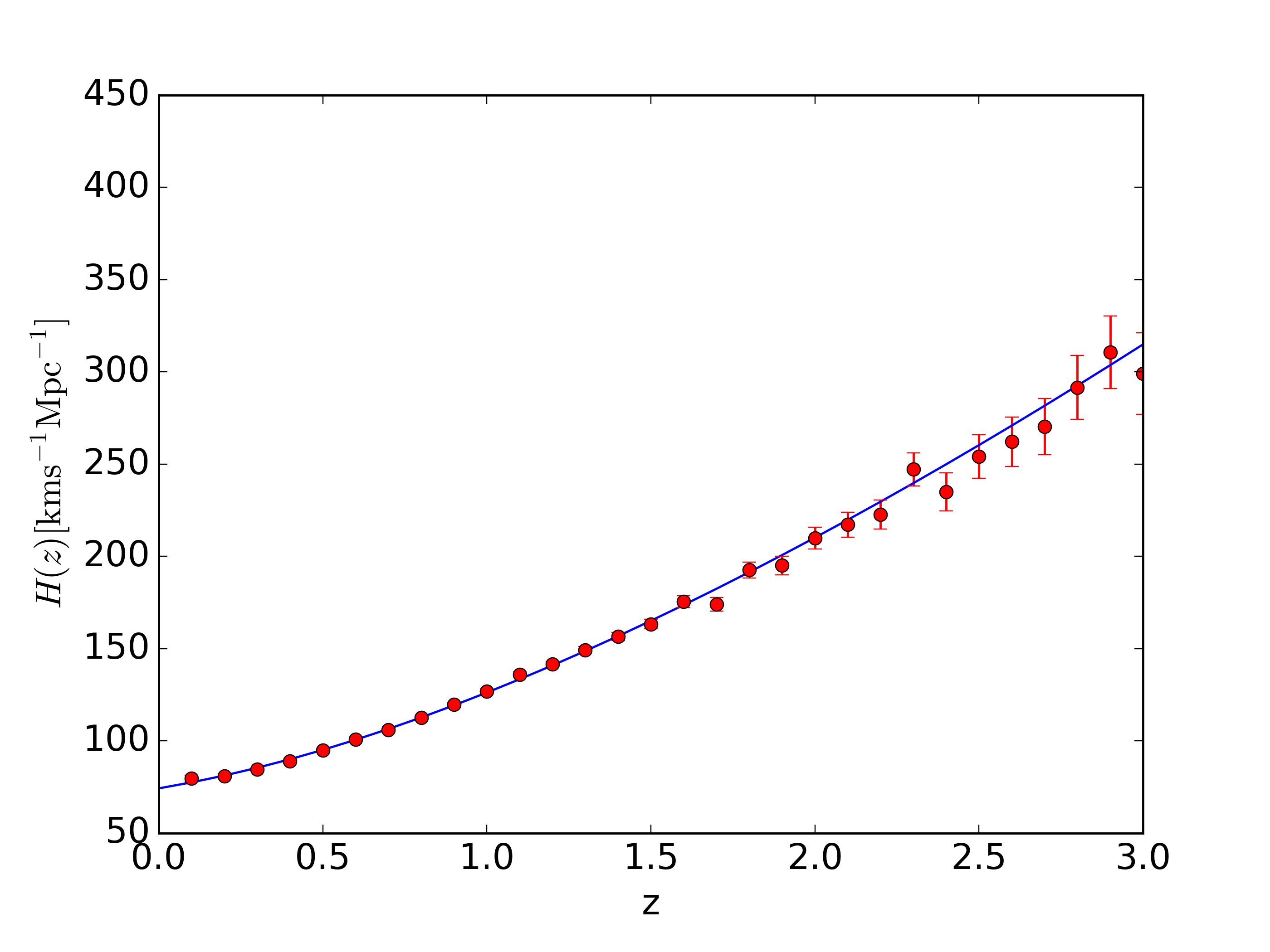}
    \caption{\label{8} Mock data for 3-year-observation. The blue curve denotes $H(z)$ value under $\Lambda$CDM, while the dots with error bars represent simulation data. }
\end{figure}

\begin{figure}[htb]
    \includegraphics[width=0.5\textwidth,bb=0 0 560 420]  {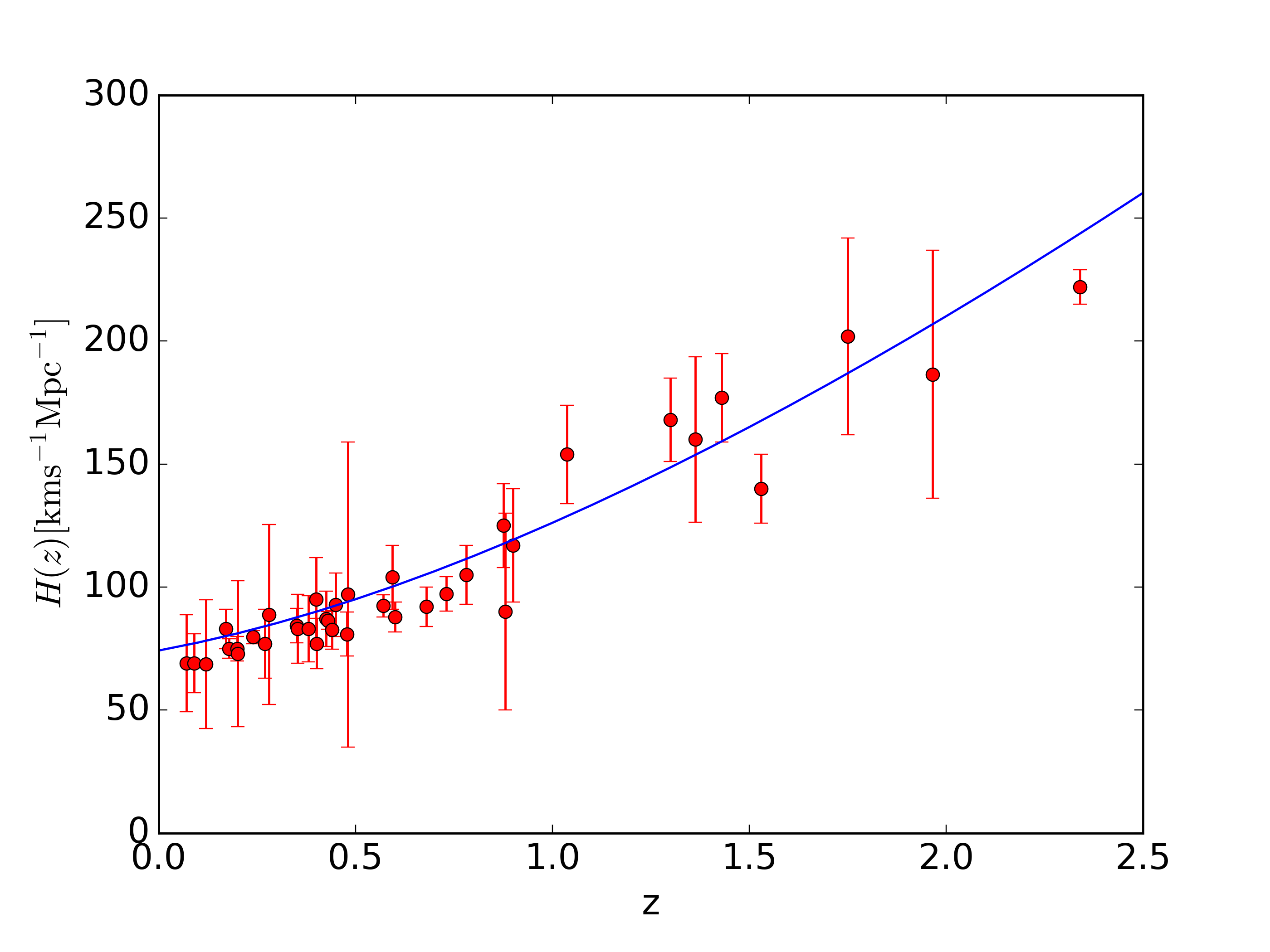}
    \caption{\label{9} 38 OHD sets. The dots with error bars represent 38 available OHD sets so far. For the purpose of illustrating, we also plot $H(z)$ value under $\Lambda$CDM, the blue curve.  }
\end{figure}

Now that we have the error information of $H(z)$, we can simulate the OHD. We follow the method Yuan Shuo \cite{2015JCAP...02..025Y} to generate mock data for $\Lambda$CDM:
\begin{equation}
    H_{sim}=H_{\Lambda CDM}+H_{drift} \text{.}
\end{equation}
We treat $H_{sim}$ as a drift, $H_{drift}$, based on the theoretical $H(z)$ value under $\Lambda$CDM, $H_{\Lambda CDM}$, caused by various errors. $H_{drift}$ is a random value under gaussian distribution, $N(0,\Delta H)$. $\Delta H$ is calculated by relative error we get in last section. Using a piece of python code, we generate our mock $H_{sim}$ data of 3-year observation at very 0.1 redshift bin, shown in Fig. \ref{8}. We get 38 OHD sets up to now. The datas were obtained by different ways from different groups \cite{2003ApJ...593..622J, 2005PhRvD..71l3001S, 2010JCAP...02..008S, 2012JCAP...07..053M,
2016JCAP...05..014M, 2014RAA....14.1221Z, 2015MNRAS.450L..16M, 2009MNRAS.399.1663G, 2012MNRAS.425..405B, 2013MNRAS.429.1514S, 2013MNRAS.431.2834X, 2015arXiv150702517M}. Fig. \ref{9} shows $H_{\Lambda CDM}$ at every redshift and the 38 OHD sets so far. As we can see, the OHD value goes up and down around the $H_{\Lambda CDM}$ at the same redshift, which justifies the validity of our simulation. Compared with our mock data, the actual OHD sets' accuracy is obviously worse.

\subsection{Forecasting}
\label{sec:velocity}

Now that we have got the 3-year-observation mock data, we can use them to forecast. Before that, we need a criteria to evaluate the constraining ability of the dataset-Figure of Merit(FoM). We can define FoM in different ways, as long as its value can reflect how tightly or loosely the data constrains parameters. Here for the convenience of our analysis, we adopt the definition by Albrecht \cite{2006astro.ph..9591A}, the reciprocal of the area enclosed by the $2\sigma$ confidence region contour, coinciding with a specially appointed confidence region under gaussian distribution.

We choose the $\Lambda$CDM  as our prior model. In such a standard $\Lambda$CDM universe with a curvature term $\Omega_{k}=1-\Omega_{m}-\Omega_{\Lambda}$, the Hubble parameter is given by
\begin{equation}
   H(z)=H_{0}E(z);  E(z)=\sqrt{\Omega_{m}(1+z)^{3}+\Omega_{\Lambda}+\Omega_{k}(1+z)^{2}} \text{.}
\end{equation}
The determination of $H_{0}$ has been carried on in different $H_{0}$ tension projects. For 7-year WMAP observation, $H_{0}=73\pm3 \rm {kms^{-1}Mpc^{-1}}$ \cite{2007ApJS..170..377S}. In this paper, we take the most recent value $H_{0}=74.2\pm3.6 \rm {kms^{-1}Mpc^{-1}}$ \cite{2009ApJ...699..539R}. The best value of $\Omega_{m}$, $\Omega_{\Lambda}$ we adopt is 0.27, 0.73 respectively, due to the coherence that they are consistent with the observations and the fact that we use these value to generate our simulation data. All the three parameters are assumed under gaussian distribution. By Bayes' theorem, the posterior probability density function of parameters is
\begin{equation}
\begin{split}
 P( \Omega_{m},\Omega_{\Lambda}| \{H_{i}\} )= &   \int P( \Omega_{m},\Omega_{\Lambda},H_{0} | \{H_{i}\} )  d H_{0} \\
  = & \int \ell( \{H_{i}\}  | \Omega_{m},\Omega_{\Lambda},H_{0} )  P(H_{0})  d H_{0}  ,
\end{split}
\end{equation}
where $\ell$ is the likelihood and $P(H_{0})$ is the prior probability density function of $H_{0}$. And the expression of $\ell$ is given by(assuming no covariance between parameters)
\begin{equation}
\begin{split}
 \ell( \{H_{i}\} | \Omega_{m},\Omega_{\Lambda},H_{0} )=&\left( \prod_{i} \dfrac{1}{\sqrt{2\pi\sigma_{i}^{2}}} \right) exp \left(-\dfrac{\chi^{2}}{2}\right), \\
 \chi^{2}=&\sum_{i}  \dfrac{[ H_{0}E(z) -H_{i}]^{2} }{\sigma_{i}^{2}}   ,
\end{split}
\end{equation}
where $\sigma_{i}$ is the uncertainty of the data $H_{i}$. And the $P(H_{0})$ is Gaussian prior, given by
\begin{equation}
P(H_{0})= \dfrac{1}{\sqrt{2\pi\sigma_{H}^{2}}} exp \left[-\dfrac{(H_{0}-\mu_{H})^{2}}  {2\sigma_{H}^{2}}\right].
\end{equation}
Then the integral can be worked out for the given Gaussian prior $P(H_{0})$. There is a point in the parameter space maximizing the probability density, $P_{max}$. Because of what we have described in last paragraph, such a point in this forecasting is \{0.27, 0.73, 74.2\}. The formula
\begin{equation}
    P=P_{max} \times exp \left(-\dfrac{\Delta \chi^{2}}{2}\right)
\end{equation}
means the contour of a given confidence region, which corresponds to the value of $\Delta \chi^{2}$. We have three parameters here, $\Omega_{m},\Omega_{\Lambda},H_{0}$. $\Delta \chi^{2}$ is statistically set to 2.3, 6.17, 11.8 respectively for $1\sigma, 2\sigma, 3\sigma$ confidence region. For a direct comparing and understanding, here we choose $2\sigma$ confidence region, namely $\Delta \chi^{2}=6.17$, when calculating FoM.

\begin{figure}[htb]
    \includegraphics[width=0.5\textwidth,bb=0 0 560 420]  {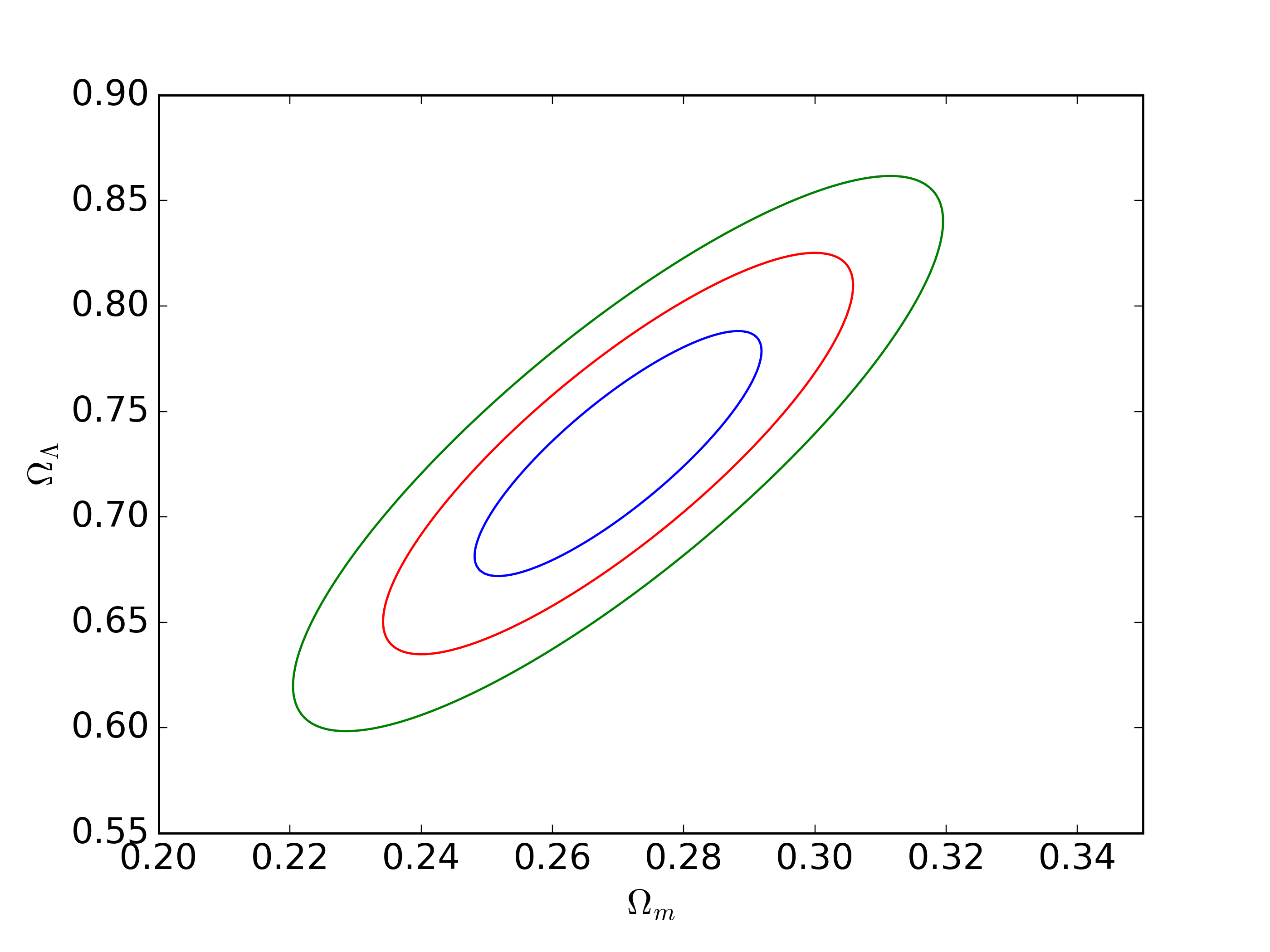}
    \caption{\label{10} Constraint on $\Omega_{m}$ and $\Omega_{\Lambda}$ for 3-year-observation. The blue, red, green curve denote $1\sigma, 2\sigma, 3\sigma$ confidence region respectively.  And the FoM of simulation data is 170.82.}
\end{figure}

To order to calculate the confidence region and FoM, we take the Fisher Matrix forecast technique \cite{2003moco.book.....D},

\begin{equation}
   F_{ij}=\dfrac{1}{2}\dfrac{\partial^{2}\chi^{2}}{\partial\theta_{i}\partial\theta_{j}},
\end{equation}
where the value of matrix elements is taken at the most-likely value of parameters. Let's denote the marginalized Fisher matrix by $\tilde{F}$, then the contour in subspace is given by
\begin{equation}
   (\Delta\theta)^{T} \tilde{F} \Delta\theta=\Delta\chi^{2}; \Delta\theta=\theta-\theta_{best-value},
\end{equation}
where $\Delta\theta$ is the deviation from the beat value of the parameters. When calculating FoM, we take $\Delta\chi^{2}$ as 6.17. The enclosed area is $\pi/ \sqrt{det(\tilde{F}/\Delta\chi^{2})}$. So FoM, the reciprocal of the area, is
\begin{equation}
   FoM=\dfrac{ \sqrt{det(\tilde{F}/\Delta\chi^{2})}}{\pi}.
\end{equation}

\begin{figure}[htb]
    \includegraphics[width=0.5\textwidth,bb=0 0 560 420]  {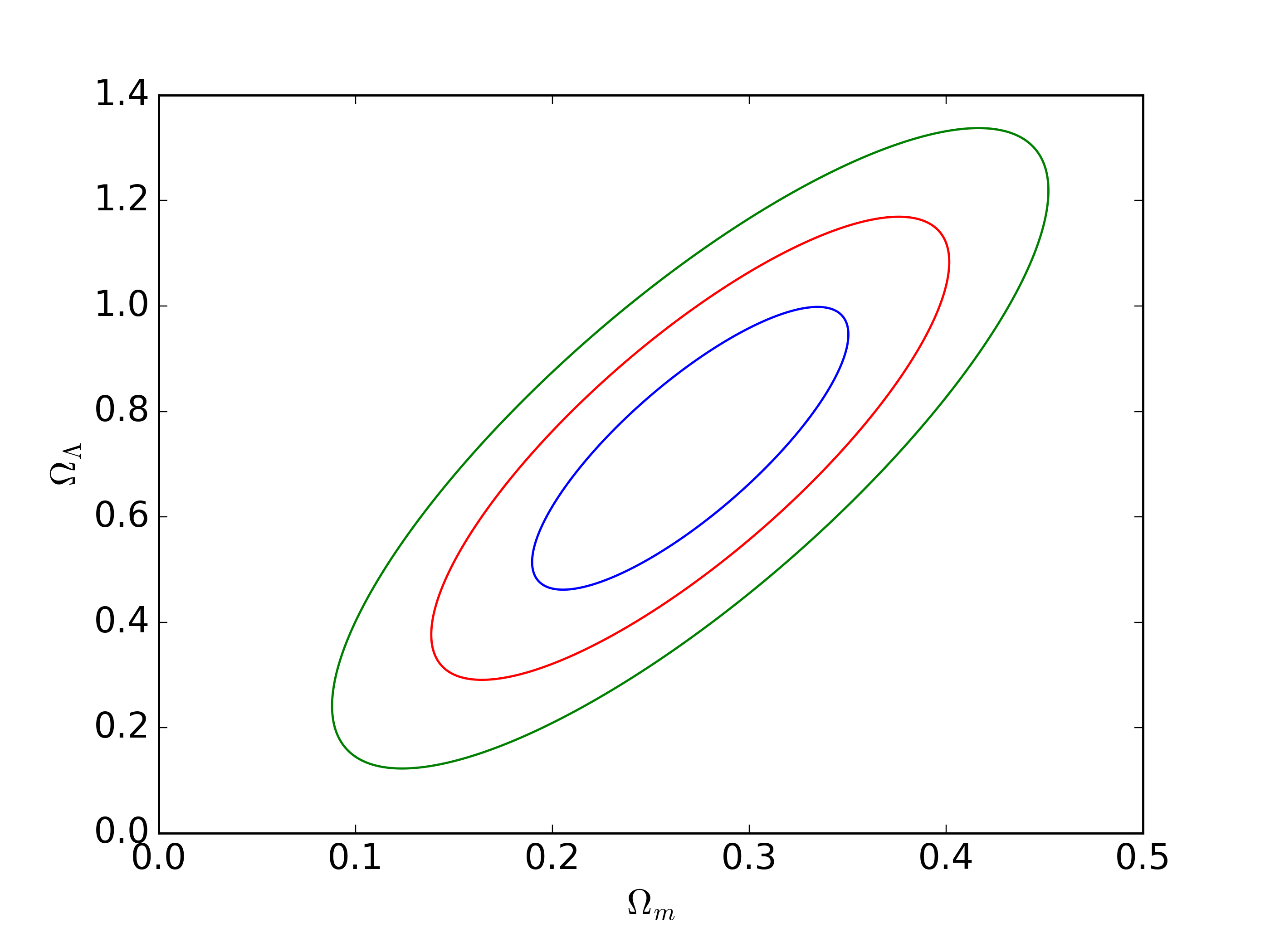}
    \caption{\label{11} Same as Fig. \ref{10}, but for 38 actual OHD sets. The FoM here is 9.3.}
\end{figure}

\begin{figure}[htb]
    \includegraphics[width=0.5\textwidth,bb=0 0 560 420]  {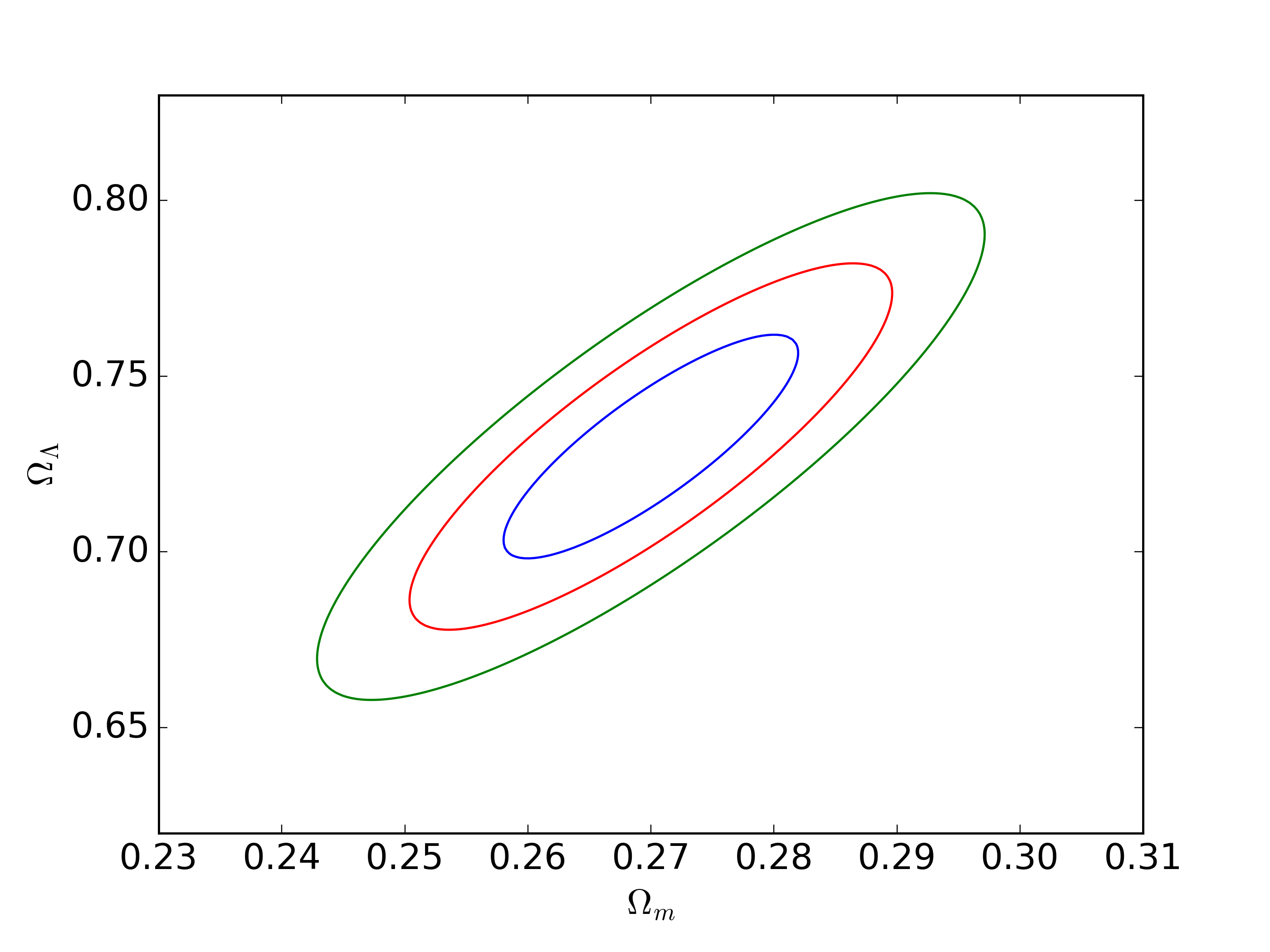}
    \caption{\label{12} Same as Fig. \ref{10}, but for 10-year-observation simulation. The FoM is 569.42.}
\end{figure}

The contour is shown in Fig. \ref{10}. As we can see, the contour is an ellipse, which is consistent with the equation of $\tilde{F}$. For a more direct and concrete comparison, we perform constraint for the 38 OHD sets. Their constraint on $\Omega_{m}$ and $\Omega_{\Lambda}$ is shown in Fig. \ref{11}. Apparently, the constraint of the mock data on parameters is much tighter, compared with that of available OHD sets, which has an significant improvement on precision cosmology. The simulation and forecast of 10-year-observation is just carried out in the same way. As Fig. \ref{12} shows, its constraint on cosmological parameters is even much tighter, implying a consequent higher FoM value.\\

For the FoM value of 3-year-observation mock data, it is about 170.82, while that of 38 OHD sets is just about 9.3. It is a remarkable improvement. For 10-year-observation mock data, the FoM has a farther improvement, reaching 569.42. We have enough reason to look forward to the excellent application of $H(z)$ data by this method.

\subsection{Nonstandard model}
\begin{figure}[htb]
    \includegraphics[width=0.5\textwidth,bb=0 0 560 420
    ]  {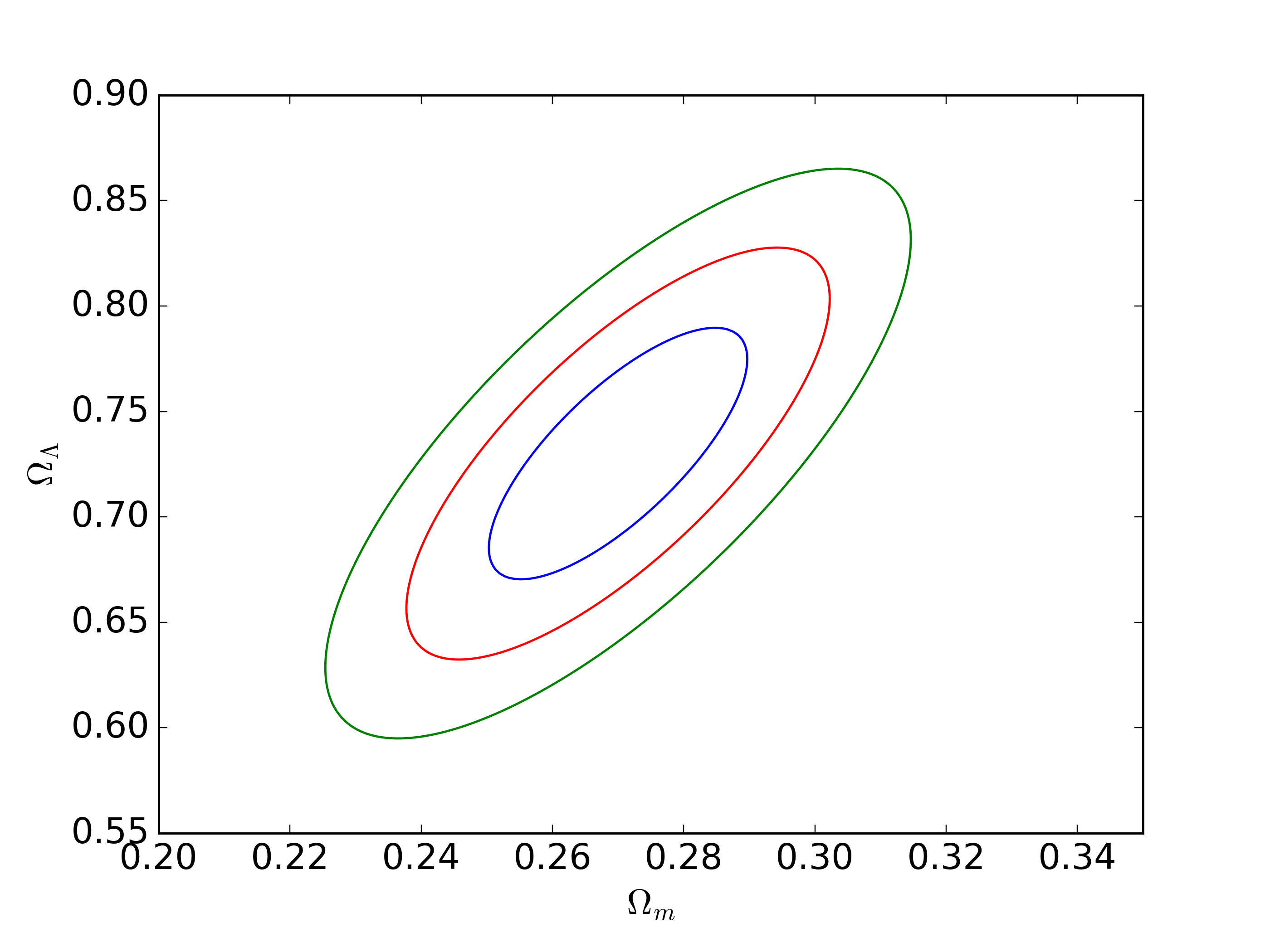}
    \caption{\label{13} The constraining of undulant universe for 3-year-observation simulation. The FoM is 153.07.}
\end{figure}

The $\Lambda$CDM universe do match the observation quite well. But it doesn't answer the question that why matter and vacuum energy should be of the same order of magnitude at this moment. Here we consider another model which can give us an answer to this problem by alternating periods of acceleration and deceleration. In undulant universe, the equation of state of the vacuum energy is an oscillatory function of state of the scale of the universe, $w(a)=-cos(In a)$. It meets the fact that $w(a=1)=-1$ in the current universe.
Then the Hubble parameter is given by:
\begin{equation}
   H(z)=H_{0}\sqrt{\Omega_{m}(1+z)^{3}+\Omega_{\Lambda}(1+z)^{3}e^{-3sin(In(1+z))}+\Omega_{k}(1+z)^{2}},
\end{equation}
where $\Omega_{\Lambda}+\Omega_{m}+\Omega_{k}=1$. The simulation and forecasting carry out just the same as above. Here we consider the case of 3-year-observation. The corresponding FoM is 153.07. And the constraint is displayed in Fig.\ref{13}

\section{Discussion and Conclusions}\label{sect:conclu}

In this paper, we mainly evaluate the quality of $H(z)$ data by GW standard sirens method of several GW detection plans, whose optimal frequency locate around the frequency window of GW from typical NS binary system. We calculate the relative error of $H(z)$ by three devices, DECIGO and ET and aLIGO. Though the sensitivity of the three devices is almost of the same order of magnitude, the H(z) error of DECIGO is quite optimistic while that of other two is far from satisfying. But it does not mean that $H(z)$ data by this method is a dead end or of no meaning, which is justified by the forecasting of DECIGO-based $H(z)$ data. If the sensitivity of aLIGO or ET is sightly improved, or just move the most sensitive frequency to a lower region, the error of $H(z)$ will be comparable that by DECIGO. To get a better $H(z)$ data, we have two ways: (1) the noise curve could be moved down, and the signal-to-noise ratio of a given sore increases, so we get more events above threshold and more "bright" events which have smaller errors; (2) the noise curve could be moved to the left, then we can see more of the inspiral which can help improve parameter estimation at fixed signal-to-noise.

Considering the absence of real $H(z)$ data by DECIGO, we simulate $H(z)$ and the data show an alluring  constraint ability on cosmological parameters. After all, we are aimed at evaluating the viability and quality of $H(z)$ data by GW standard sirens method, not putting the method into actual operation. We find that, under $\Lambda$CDM universe, the FoM of mock data shows a huge improvement compared with that of 38 actual OHD sets. For contrast, the FoM is 9.3 for 38 OHD sets, 170.82 for 3-year-observation, 569.42 for 10-year-observation respectively. The advantage of $H(z)$ is that it is the direct measurement of the expansion history, so $H(z)$ can be powerful in constraining nonstandard universe. Besides the standard model, we also explore its ability when applied to undulant universe. $H(z)$ by DECIGO still shows a excellent constraining ability and a comparably excellent result. For 3-year-observation simulation, the FoM is 153.07. The tight constraint of mock data and the FoM of the corresponding contour indicate a bright future of measuring $H(z)$ data by this method.

\begin{figure}[htb]
    \includegraphics[width=9cm,bb=0 0 560 420]  {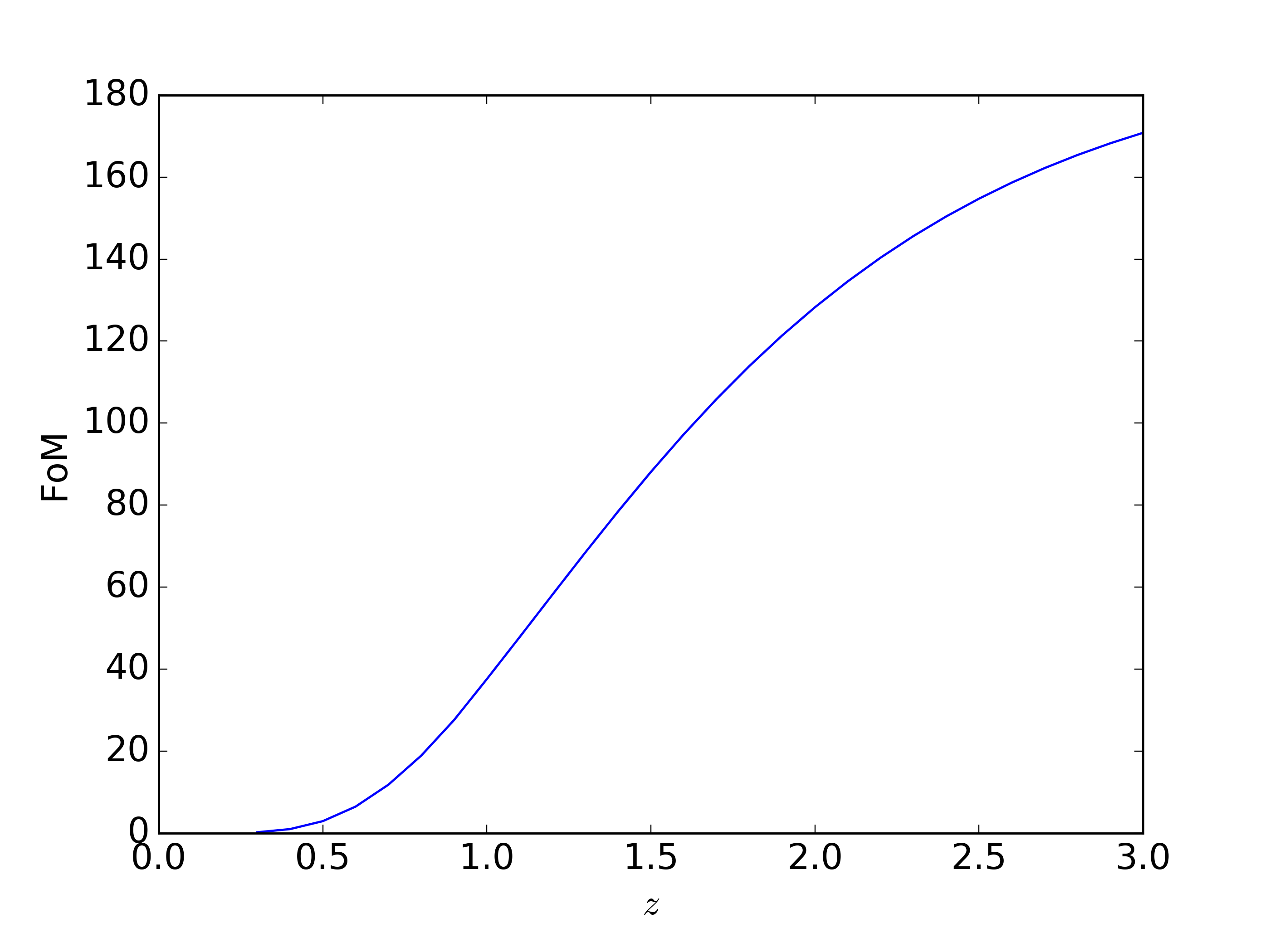}
    \caption{\label{14} The FoM variation with $z$. This Figure shows the corresponding value of FoM under the condition that if the detecting range reach redshift $z$. The result is based on 3-year-observation}
\end{figure}

To extract as much physical information as possible, all the known sources of error should be quantified. Apart from the three error mentioned above, there is another kind of errors, the calibration error. In GW detection, the response function is used to convert the electronic output of a GW detector into the measured GW signal. The calibration error is produced on the experimentally measurement of the response function \cite{2009PhRvD..80d2005L}. The calibration error in the response function degrades the ability to measure the physical properties of the GW source. Thus it is meaningful to investigate the calibration error. Lee Lindblom \cite{2009PhRvD..80d2005L} derived the optimal calibration accuracy: the lower accuracy level would reduce the quantity and the quality of the scientific information extracted from the data, and the higher accuracy would be made irrelevant by the intrinsic noise level of the detector. And S. Vitate at al. \cite{2012PhRvD..85f4034V} also investigated the effect introduced by calibration error based on the estimates obtained during LIGO's fifth and VIRGO's third science runs. They found that the calibration error would slightly damage the parameter estimate in GW data analysis. But the calibration-introduced system has a better ability in locating the source, facilitating the EM counterpart  detecting. Considering the damage caused by calibration error is relatively small and it¡¯s hard to quantify the calibration error, we ignore it in this work. We expect future study can give us more precise estimate.

It is well worth to point out that at current stage, the detecting of electromagnetic counterpart is still a problem. Currently, we are not able to make a good evaluation and conclusion about it. Finding the EM counterpart in the GW event is crucial for GW astronomy, which can reveal the process and interaction during the merger process \cite{2016MNRAS.459..121C}. Mwtzger et al. \cite{2012ApJ...746...48M} showed that the transient EM counterpart can possibly occur within a few seconds after the binary merger. And a lot of theoretical and experimental progress have been achieved \cite{2012A&A...541A.155A, 2012SPIE.8448E..0QS}.

Also there are some recent development in astronomical and computing technologies. During the proposal and test of a number of low-latency GW trigger-generation pipelines, the pipeline has been improved and capable of generating event triggers within minutes upon the arrival of a detectable signal \cite{2012A&A...541A.155A, 2010CQGra..27s4013B, 2012ApJ...748..136C, 2012PhRvD..86b4012H}. More and more detectors to be constructed can form a network, rendering it likely to improve the localization efficiency \cite{2011CQGra..28j5021F, 2011PhRvD..84j4020V, 2010PhRvD..81l4048Z}. Some methods have been proposed to identifying GW source for a large sky error \cite{2013ApJ...767..124N, 2015ApJ...814...25C}. Considering the fact that the early detector network¡¯s error in GW localization will be of order $200 deg^{2}$ \cite{2014ApJ...795..105S}, such method would improve the feasibility of EM detector a lot. The devices that aim at facilitating the prompt EM detection mainly focus on high energy region and the optical region, while radio region is also a good candidate \cite{2016MNRAS.459..121C}.  By next decade, the Large Synoptic Survey Telescope(LSST), will be in its sky survey. It will bring us great hope to find the prompt EM counterpart. Such EM detection demand multi-wavelength programs by sensitive telescope capable of covering large areas on the sky, and a strong synergy exists between LSST and radio survey in identifying the EM counterpart at both optic and radio wavelengths, and the information from both wavelengths about the physics of the post-merger will be complementary \cite{2014PASP..126..196L}. Here we stress the evaluation of $H(z)$ by standard sirens, not the exact technical details. Another problem is that the detecting range for NS binary system is just 300Mpc now \cite{2011CQGra..28l5023S}. This range is much smaller than what we assumed above. We explore how the FoM changes with the variation of detecting range, which is shown in Fig. \ref{14}. The FoM can be comparable with that of 38 OHD sets at $z=0.7$. For 10-year-observation, this critical value would be smaller.In other words, if we launch a 10-year-observation, even if the detecting range is just z=0.7, we can do much better than 38 OHD sets. The reason why the limited data can produce such good effect lies on the fact that the GW standard sirens can measure low-$z$ $H(z)$ with excellent accuracy. This demonstrates that even if we could not detect the high-$z$ $H(z)$ data by GW standard sirens method, the low-$z$ data can still be valuable and powerful. In the further, if we want to measure high-$z$ $H(z)$ by this way, some improvement, probably a lower strain noise, is necessary. But it is undoubtable that the $H(z)$ by this method is of great power and potential.

\section*{acknowledgements}

We sincerely thank the referee for his/her useful responses, which help us greatly improve our manuscript. This work was supported by National Key R$\&$D Program of China (2017YFA0402600), the National Science Foundation of China (Grants No. 11573006, 11929301, 61802428) and Shandong Provincial Natural Science Foundation of China(Grant No. ZR2019MA059).

\clearpage
\bibliographystyle{spphys}
\bibliography{Omh2}
\end{document}